\documentstyle[12pt,aasms4]{article}

\slugcomment{}

\lefthead{Miyata et al.}
\righthead{Radial profile the Cygnus Loop}

\doublespace
\newcommand{\choi}{\hspace{1mm}}
\newcommand{\chobi}{\hspace{0.5mm}}

\newcommand{\kara}{{\mbox{\rm\choi $\sim$}}}
\newcommand{\kaijo}[2]{\mbox{$\rm{\choi{#1}^{\scriptscriptstyle #2}}$}}

\newcommand{\Msun}{{\mbox{$\chobi M_{\odot}$}}}

\begin{document}

\title{The Radial Structure of the Cygnus Loop Supernova Remnant \\
  --- Possible evidence of a cavity explosion ---}
\author{Emi Miyata\altaffilmark{1} and Hiroshi Tsunemi\altaffilmark{1}}
\affil{Department of Earth and Space Science,
	Graduate School of Science, Osaka University \\
	 1-1, Machikaneyama, Toyonaka, Osaka, 560-0043, Japan}

\altaffiltext{1}{CREST, Japan Science and Technology
	Corporation (JST), 2-1-6 Sengen, Tsukuba, Ibaraki 305-0047}

\begin{abstract}

 We observed the North-East (NE) Limb toward the center region of the
 Cygnus Loop with the ASCA Observatory. In our previous paper (Miyata et
 al.  1994), we analyzed the data obtained with the X-ray CCD cameras
 (SISs).  We found a radial variation of electron temperature ($kT_{\rm
 e}$) and ionization timescale (log($\tau$)) whereas no variation could
 be found for the abundances of heavy elements.  In this paper, we
 re-analyzed the same data set and new observations with the latest
 calibration files.  Then we constructed the precise spatial variations
 of $kT_{\rm e}$, log($\tau$), and abundances of O, Ne, Mg, Si, and Fe
 over the field of view (FOV).  We found a spatial variation not only in
 $kT_{\rm e}$ and in log($\tau$) but also in most of heavy elements. As
 described in Miyata et al. (1994), values of $kT_{\rm e}$ increase and
 those of log($\tau$) decrease toward the inner region. We found that
 the abundance of heavy elements increases toward the inner region. The
 radial profiles of O, Ne, and Fe show clear jump structures at a radius
 of 0.9 $R_{\rm s}$, where $R_{\rm s}$ is the shock radius. Outside of
 0.9 $R_{\rm s}$, abundances of all elements are constant. On the
 contrary, inside of 0.9 $R_{\rm s}$, abundances of these elements are
 20--30 \% larger than those obtained outside of 0.9 $R_{\rm s}$.  The
 radial profile of $kT_{\rm e}$ also shows the jump structure at 0.9
 $R_{\rm s}$.  This means that the hot and metal rich plasma fills the
 volume inside of 0.9 $R_{\rm s}$. We concluded that this jump structure
 was the possible evidence for the pre-existing cavity produced by the
 precursor.  If the ejecta fills inside of 0.9 $R_{\rm s}$, the total
 mass of the ejecta was roughly 4\Msun. We then estimated the
 main-sequence mass to be roughly 15\Msun, which supports the massive
 star in origin of the Cygnus Loop supernova remnant and the existence
 of a pre-existing cavity.

\end{abstract}


\pagebreak

 \section{Introduction}

The X-ray emission of the shell-type supernova remnants (SNRs) is mainly
due to the thermal emission from the shock heated plasma. The standard
model of the dynamical evolution of SNRs shows that the reverse shock
wave propagates into the ejected material and the fore shock wave
expands into the ambient medium (McKee 1974). The shocked material is
separated at the contact discontinuity between the ejecta and the
interstellar medium (ISM). Many theoretical calculations have been
performed to follow the dynamical evolution of SNRs (e.g. Mansfield \&
Salpeter 1974, Chevalier 1982).

Just after being engulfed by the blast wave, the shocked matter is
accelerated up to 3/4 $V_{\rm s}$, where $V_{\rm s}$ is the shock
velocity, due to the mass conservation. The kinetic energy of electron
gas, 1/2 $m_{\rm e}$ (3/4$V_{\rm s}$)$^2$, is roughly 2000 times smaller
than that of ion gas, 1/2 $m_{\rm i}$ (3/4$V_{\rm s}$)$^2$, where $m_{\rm
e}$ and $m_{\rm i}$ are masses of electron and ion, respectively.
Shklovskii (1962) proposed that the equipartition between the ion gas
and the electron gas could be established only by Coulomb collisions. In
this case, the timescale for the equipartition is longer than the age of 
SNRs (Itoh 1978). The model to consider the deviation from the thermal
equilibrium is called a two-fluid model. On the other hand, McKee
(1974) proposed that collisionless shocks with high Mach number could
equilibrate $T_{\rm e}$ and $T_{\rm i}$ by plasma instabilities or
turbulence (one-fluid model), where $T_{\rm e}$ and $T_{\rm i}$ are
the temperatures of electron gas and ion gas, respectively.

The previous X-ray observations (Aschenbach 1985; Bleeker 1990) show
that the $kT_{\rm e}$ of young SNRs indicates that the electron gas is
heated up to several keV, which is much higher than that expected from
the non-equipartition model. Bleeker (1990) suggested that $T_{\rm e}$
$\simeq$ $T_{\rm i}$ was not grossly in error. On the other hand, based
on the analysis of the optical observations of Balmer-dominant SNRs,
Smith et al.  (1991) showed that the ratios of broad to narrow H$\alpha$
intensity corresponded to neither the one-fluid nor the two-fluid
model. The results of plasma measurements near the Earth bow shock
support the two-fluid model (Montgomery et al.  1970).

The Cygnus Loop is one of the best-studied SNRs in radio,
infrared, optical, and X-ray wavelengths. Its high surface brightness
and large apparent size make it an ideal object for the study of the
various shock heating conditions and spatially-resolving spectroscopy in
detail. Various optical observations of the Cygnus Loop support a cavity
explosion model which was initially proposed by McCray \& Snow
(1979). Hester \& Cox (1986) and Hester et al. (1994) investigated the
optical emission as well as the X-ray emission at bright shell regions
of the Cygnus Loop and concluded that the bright X-ray emission
originated not from evaporative clouds but from locally high density
regions, suggesting the presence of large clouds around the Cygnus
Loop. Shull \& Hippelein (1991) measured proper motions at 39 locations
within the Cygnus Loop and found an asymmetric expansion of the
shell. They also concluded the existence of a pre-existing cavity wall
around the Cygnus Loop. This model is supported by observations of
coronal iron line emission (Teske 1990) and by observations with IRAS
(Braun \& Strom 1986).

Charles et al. (1985) observed the western edge of the Cygnus Loop with
the Einstein Observatory. They found that $kT_{\rm e}$ increased toward
the center and that the emission measure was proportional to the inverse
square of $kT_{\rm e}$. They concluded that the bright X-ray emission
was due to cloud evaporation and suggested the presence of a
pre-existing cavity wall.  Miyata et al.  (1998) found the mass of the
progenitor star of the Cygnus Loop to be $\simeq$ 25\Msun. This result
also supports the presence of the pre-existing cavity.

In this {\it paper}, we present the radial profile from the NE limb
toward the central region. Miyata et al.  (1994; hereafter MTPK)
observed the NE portion of the Cygnus Loop with the ASCA Observatory.
They obtained moderate resolution X-ray spectra with the X-ray CCD
camera, SIS. They determined both $kT_{\rm e}$ and the abundances of
heavy elements.  They found no spatial variations of the abundances of
heavy elements.  Data set we used for the NE limb was that of the same
as MTPK.

\section{Observations and Data Corrections}

The NE limb (reg-1) and the inner region (reg-2) of the Cygnus Loop were
observed with ASCA for \kara 8 and \kara 10~ks on Apr. 20 and
Dec. 17-18, 1993, respectively. The ASCA Observatory, the fourth
Japanese X-ray satellite (Tanaka, Inoue, \& Holt 1994), is equipped with
four X-ray telescopes (XRT; Serlemitsos et al.  1995) that
simultaneously feed four focal plane instruments.  The data reported
here were obtained from the two Solid-state Imaging Spectrometers (SIS0
and SIS1; Yamashita et al. 1997) which have a FOV of 22\arcmin $\times$
22\arcmin\choi and an energy resolution of approximately 60 eV full
width at half maximum below 1~keV.  The telescopes have a point spread
function with a half power diameter (HPD) of about 3\arcmin.

In this paper, we re-analyzed the SIS data obtained at the NE limb using
the latest software and calibration files available at the time of this
writing.  We employed {\tt Ftools} ver 3.5 and {\tt Xselect} ver 1.3 to
extract our data sets. We excluded all the data taken at elevation
angles below 5$^\circ$ from the night earth rim and 50$^\circ$ (for SIS)
from the day earth rim, a geomagnetic cutoff rigidity lower than 6 GeV
c$^{-1}$, and the region of the South Atlantic Anomaly.  After screening
with above criteria, we manually removed the time region where we could
see a sudden change in light curves of the corner pixels of X-ray
events.  Then, we removed the hot and flickering pixels and corrected
CTI, DFE, and Echo effects (T. Dotani et al. 1995, ASCA Letter News 3,
25; T. Dotani et al. 1997, ASCA Letter News 5, 14) in our data sets by
using {\tt sispi} and {\tt faint} commands with the calibration file of
{\tt sisph2pi\_290296.fits}. We considered the time-dependent effects of
CTI and Echo whereas MTPK did not. Since these effects could be
corrected only for the Faint-mode data, we focused only on the
Faint-mode data for the spectral analysis.  The observing times of the
Faint-mode data for reg-1 and reg-2 were $\sim$ 7.3 and $\sim$ 4.6 ks
after screening the data.  We subtracted a blank-sky spectrum (North
Ecliptic Pole and Lynx field regions) as the background, since we
estimated the contribution of the Galactic X-ray background to be
negligibly small (Koyama et al. 1986). The count rates for reg-1 and
reg-2 in our FOV were $\sim$ 13 and $\sim$ 6.8 c ${\rm s^{-1}}$ /SIS.
We used {\tt xanadu} ver 8.5 for the spectral analysis performed in this
paper.

\begin{figure}[htbp]
  \centering
	\plotone{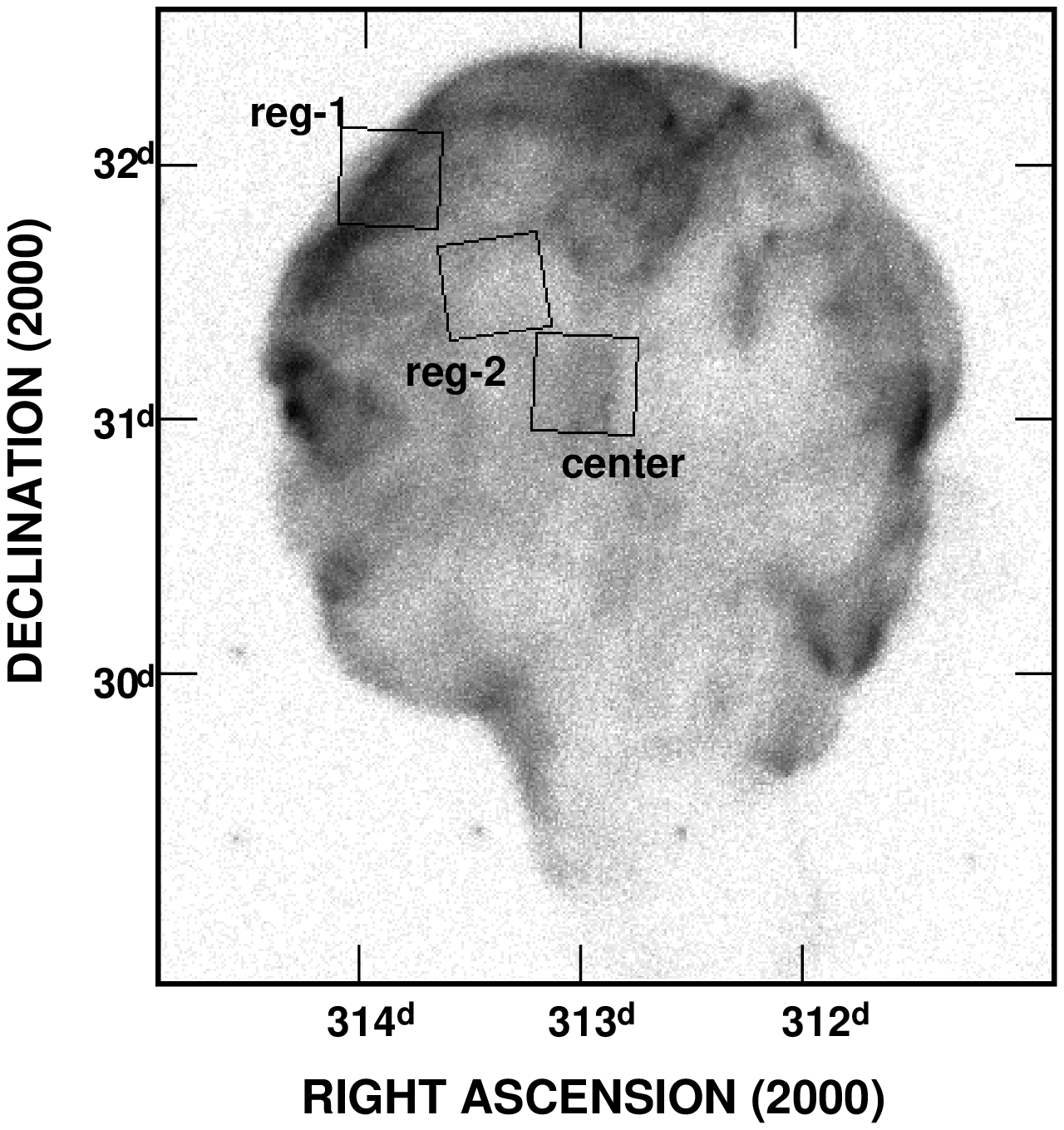}
	\caption{X-ray surface brightness map of the Cygnus Loop
	obtained by the ROSAT all-sky survey (Aschenbach 1994).  The
	black squares show the FOVs of reg-1, reg-2, and the center
	portion as observed with the ASCA Observatory.}
	\label{fig:ana:limb:image}
\end{figure}

Figure~\ref{fig:ana:limb:image} shows the location of our SIS FOVs
superimposed on the X-ray surface brightness map of the Cygnus Loop
(Aschenbach 1994) with black squares.  The location of the observation
by Miyata et al.  (1998) is also shown as `center' in this figure.

\section{Spectral Analysis}

\begin{figure}[htbp]
  \centering
	\plotone{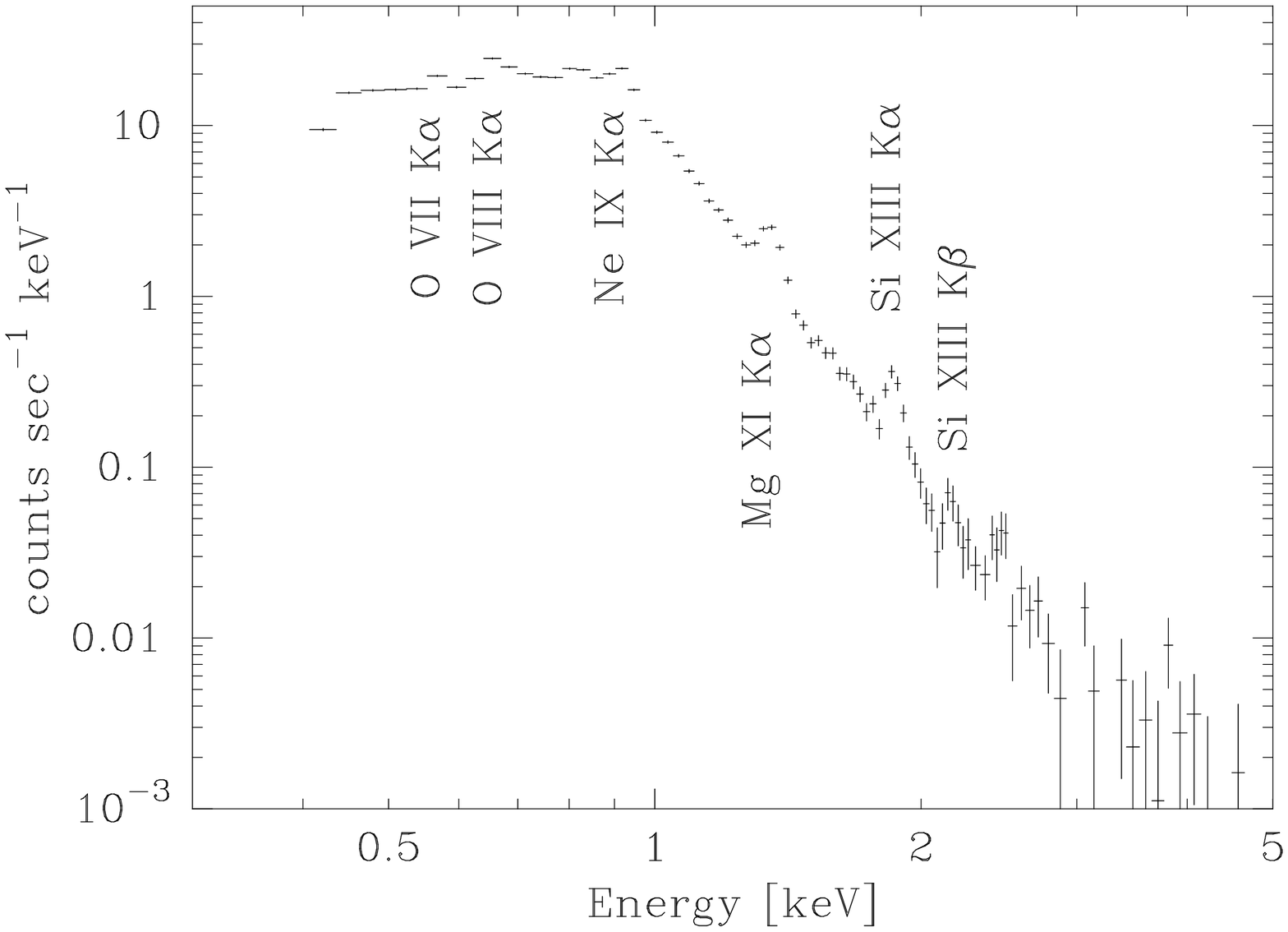}

	\plotone{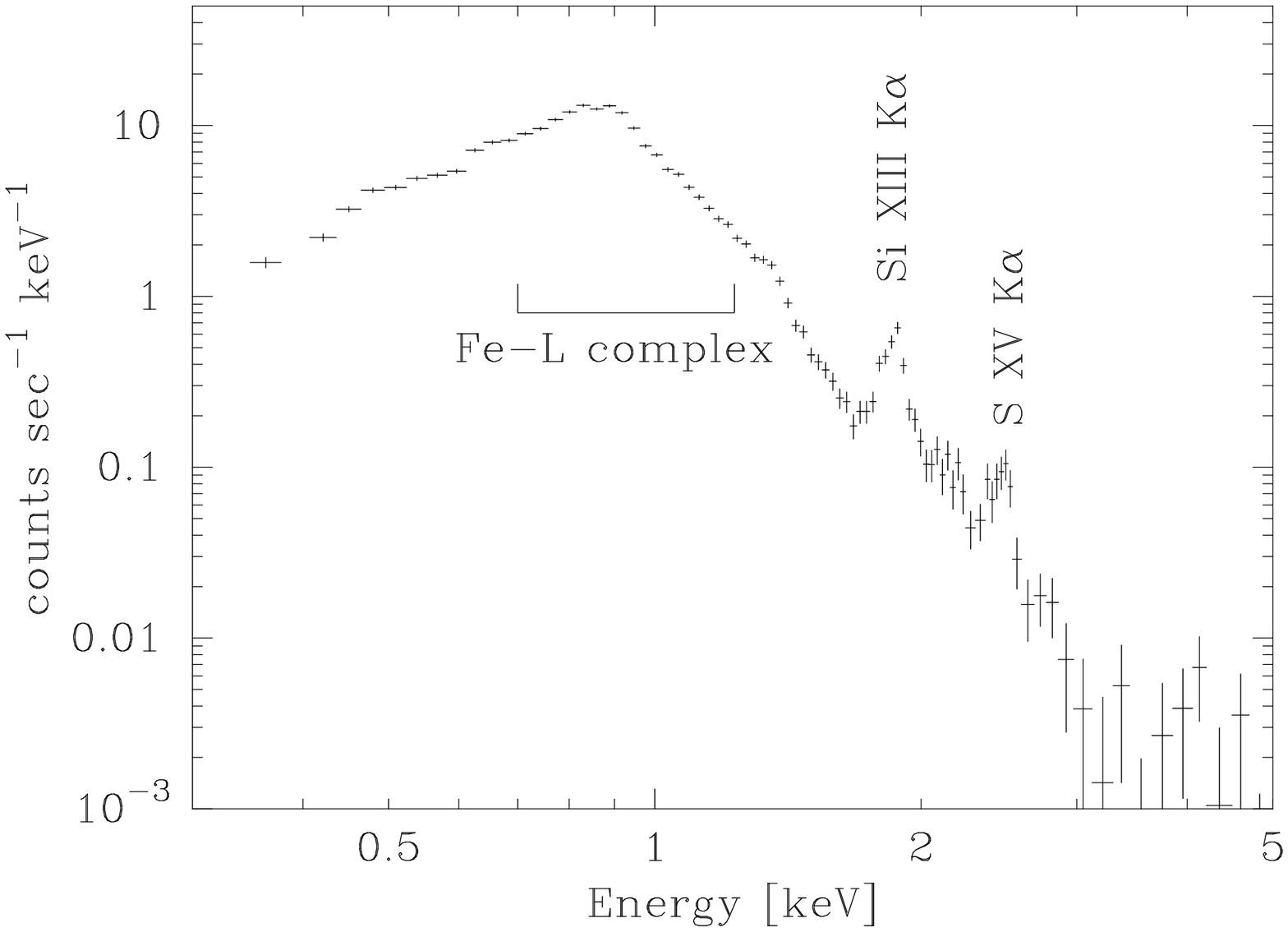}
	\caption{Spatially integrated spectra both at reg-1 and reg-2.
 Line identifications are also shown.}
	\label{fig:ana:limb:whole}
\end{figure}

Figure~\ref{fig:ana:limb:whole} shows the spatially integrated spectra
both of reg-1 and of reg-2 after subtracting background. We clearly see
K emission lines from O, Ne, Mg, Si, and S and L emission lines from Fe
in reg-1.  In reg-2, Si and S emission lines are strong whereas the
other emission lines which appeared in reg-1 are not clearly seen.  In
the following fit procedures, we fixed the neutral H column to be $4
\times 10^{20} \ {\rm cm}^{-2}$ (Inoue et al.  1979; Kahn et al.  1980).

\subsection{Spectral Variation over the FOV}

We investigated the spatial variations of the plasma structure over the
FOV. We divided our FOV into 3\arcmin\choi squares each of which
corresponds to the HPD of the XRT (Serlemitsos et al. 1995).
Furthermore, we arranged the small squares such that each square was
overlapping by 2\arcmin\choi with adjacent squares, resulting 400 squares
for each region.  Since the optical axes of SIS0 and SIS1 were different
from each other, we employed the {\tt ascatool} library to adjust the
optical axes.

Then, we applied the single component non-equilibrium ionization (NEI)
model coded by K. Masai (Masai 1984; Masai 1994) to each spectrum. Free
parameters of the fitting were $kT_{\rm e}$, log($\tau$), abundances of
O, Ne, Mg, Si, Fe, and Ni, and emission measure (EM [\kaijo{cm}{-6}
pc]). Values of reduced $\chi^2$ scattered from 1 to 2.

\begin{figure}[htbp]
 \centering
 \epsscale{.8}
	\plotone{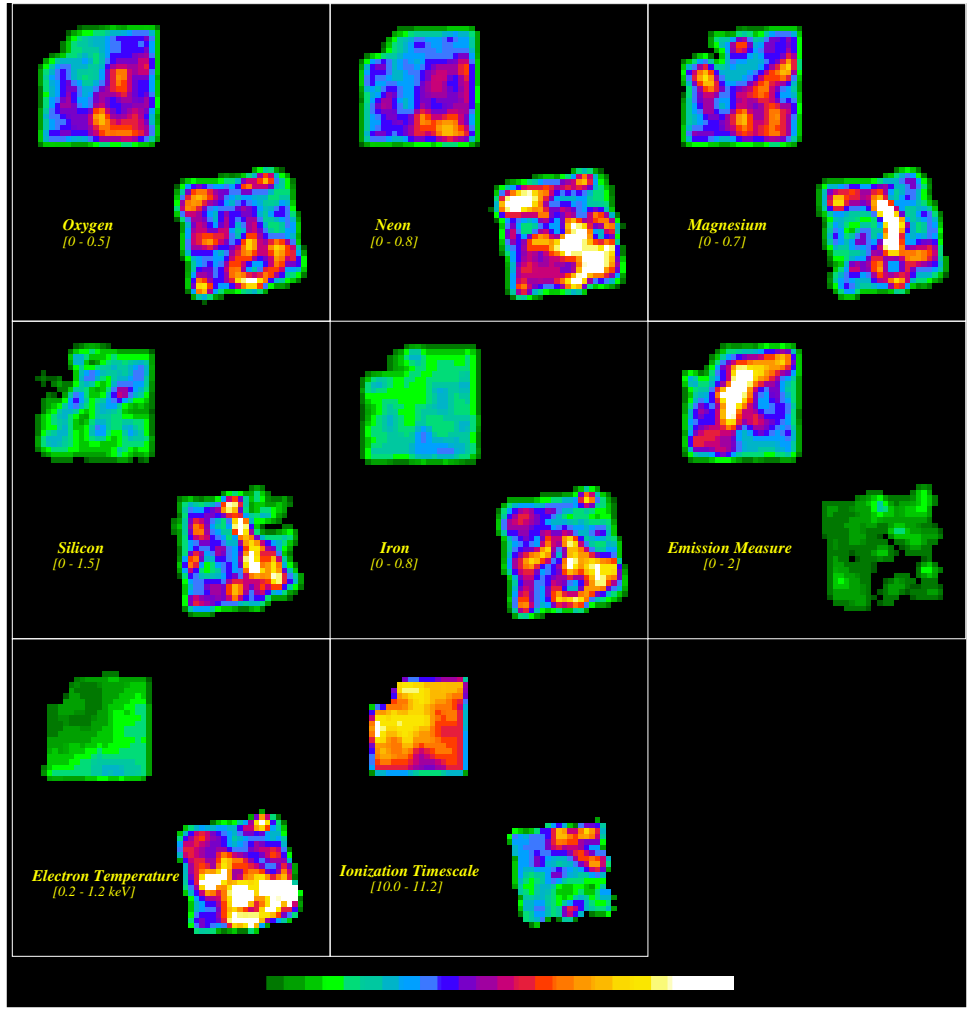}
 \caption{Spatial distribution of abundances of O, Ne, Mg, Si,
 and Fe and EM, $kT_{\rm e}$, and log($\tau$). Abundances,
 $kT_{\rm e}$, and EM
 are shown in cosmic abundance, in keV, and in
 ${\rm cm}^{-6}$pc, respectively. Each image was smoothed with
 a Gaussian function of FWHM=3$^\prime$.}
 \label{fig:ana:limb:map}
\end{figure}

Figure~\ref{fig:ana:limb:map} shows the spatial variations of abundances
of O, Ne, Mg, Si, and Fe relative to the cosmic abundance, EM, $kT_{\rm
e}$, and log($\tau$). Filamentary-like structures are found in Si and Mg
maps. However, we focused on the global structure of distribution of
heavy elements. Abundances of heavy elements increase toward the inner
region even in the cases of Si and Mg maps. Abundance variations can be
found even inside reg-1. Except Si, abundances are smaller than the
cosmic values. Obtained values of emission measure are anti-correlated
with those of $kT_{\rm e}$ whereas they are well correlated with those
of log($\tau$). These facts were already found in reg-1 by MTPK.

\subsection{Radial Distribution of the Plasma Structure}

\begin{figure}[htbp]
 \centering
	\plotone{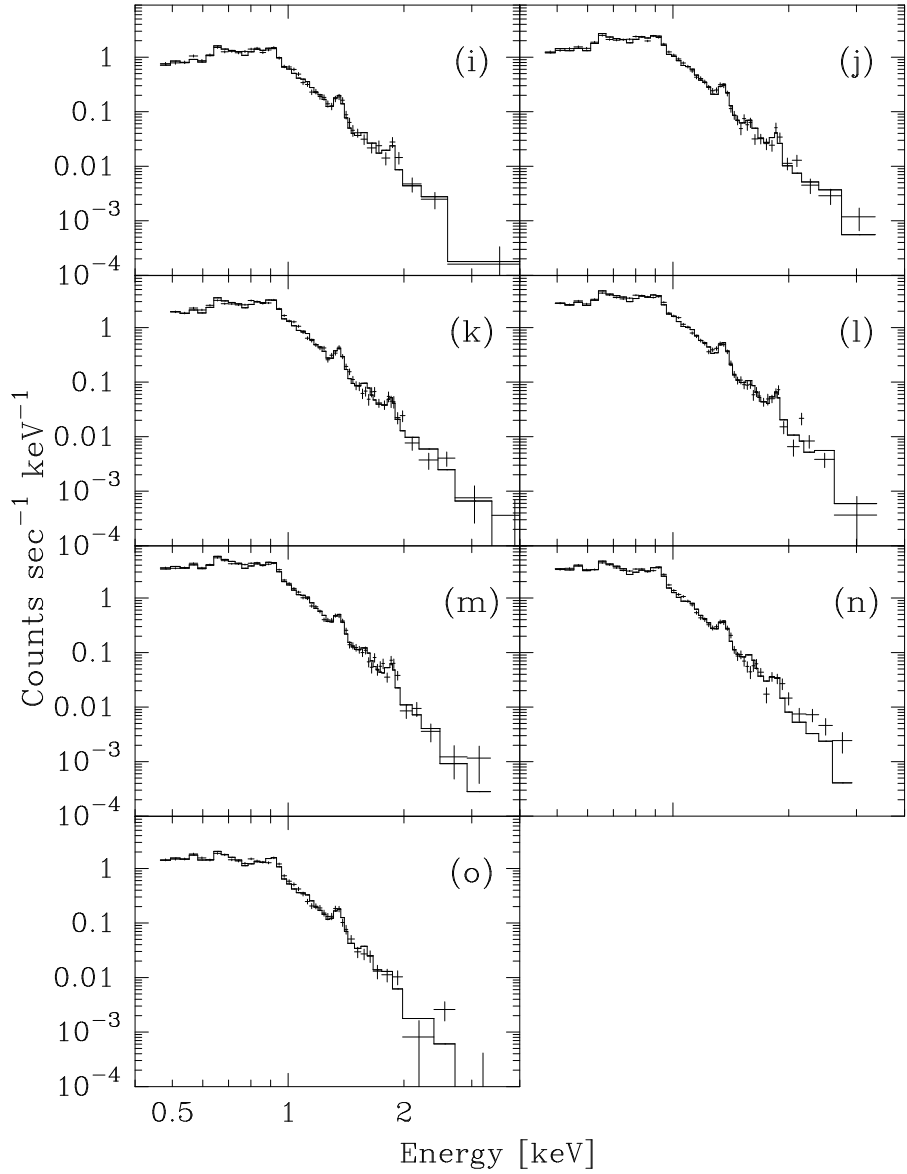}
 \caption{Best fit curves of the Masai model in reg-1}
 \label{fig:ana:limb:masai}
\end{figure}

\begin{figure}[htbp]
 \centering
	\plotone{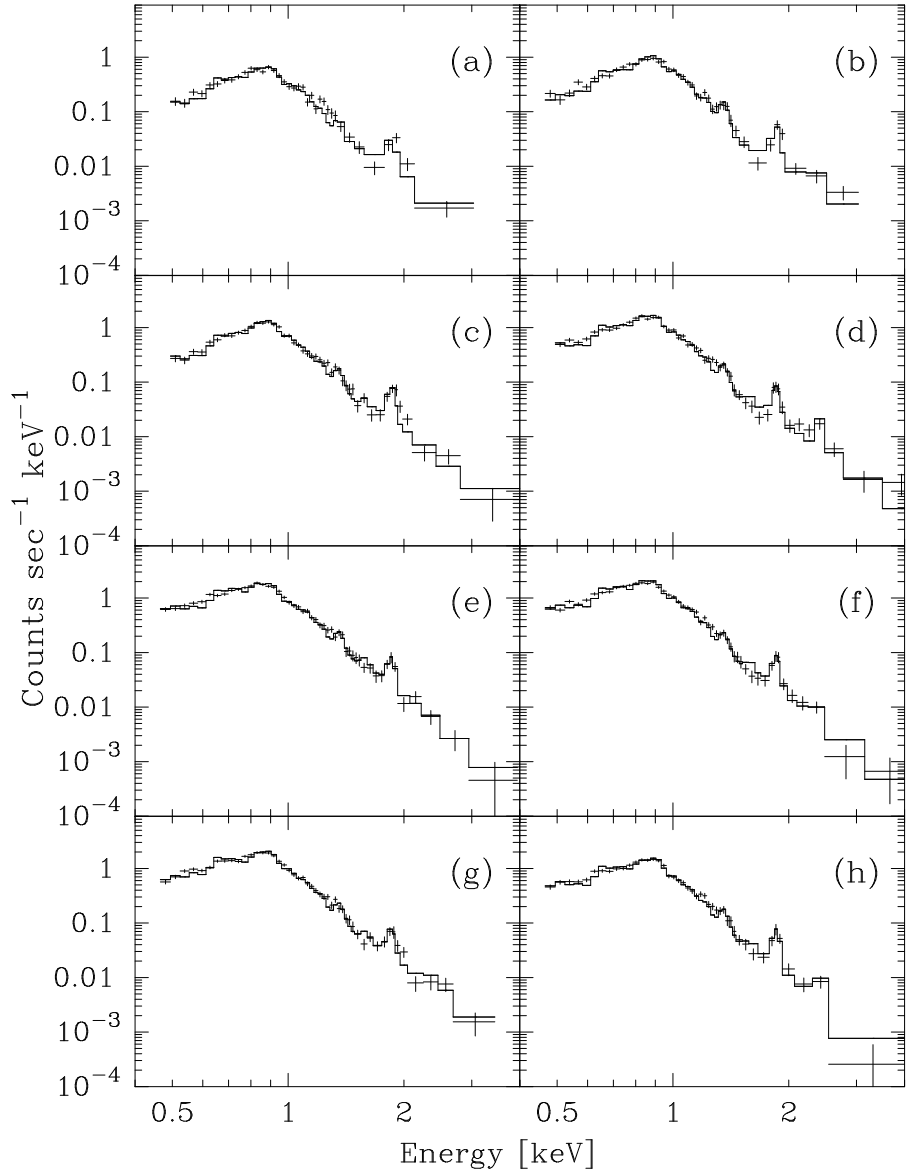}
 \caption{Same as figure 4 but for reg-2}
 \label{fig:ana:limb2:masai}
\end{figure}

As shown in figure~\ref{fig:ana:limb:map}, we found that there were
structures in the radial direction rather than the azimuthal
direction in the abundance maps as well as in other physical parameters.
Therefore, we divided our FOV into 7 and 8 annular sectors for reg-1 and
reg-2 to investigate the radial distribution of plasma structure as
performed by MTPK.  Each sector was separated by \kara 3\arcmin.  The
spectrum extracted from each region is shown in
figures~\ref{fig:ana:limb:masai}--\ref{fig:ana:limb2:masai}. All spectra
have been background subtracted properly.  We noticed that slopes in the
energy range of $0.5-0.9$ keV generally change from (i) to (o) in reg-1.
This indicated that values of $kT_{\rm e}$ decreased from the inner
region to the outer region. We also noticed that the line intensity
ratios of \ion{O}{8} to \ion{O}{7} increased toward the center,
suggesting higher $kT_{\rm e}$ or lower log($\tau$) at the inner
region. In reg-2, Si emission line is generally stronger than that in
reg-1. The continuum emission in reg-2 extends to higher energy than
that in reg-1, suggesting higher $kT_{\rm e}$ in reg-2 compared with
that in reg-1.

We again applied the single component Masai model to the 15
spectra. Free parameters are the same as those used in the previous
section.  Single component Masai model gave us reasonable fits to all
regions. Best fit curves are shown in
figures~\ref{fig:ana:limb:masai}--\ref{fig:ana:limb2:masai} by solid
lines.  The obtained $kT_{\rm e}$, log($\tau$), and EM are summarized in
figure~\ref{fig:ana:limb:kne}. We, here, assumed the shock radius,
$R_{\rm s}$, to be 84\arcmin (Ku et al. 1984; hereafter KKPL).

\subsubsection{$N_{\rm H}$ Uncertainties}

Leahy et al. (1994) found that $N_{\rm H}$ was not uniform across the
Cygnus Loop and increased from west-southwest to east-northeast.  Based
on their results, $N_{\rm H}$ changed from the center portion ($\approx
3.0 \times 10^{20}$cm$^{-2}$) toward the eastern region ($\approx 5.0
\times 10^{20}$cm$^{-2}$).  The angular resolution of their observation
is not high enough to compare with our data sets but there are no other
results for $N_{\rm H}$ variations.  Therefore, we fixed $N_{\rm H}$ to
be 2.0$\times 10^{20}$ and 6.0$\times 10^{20}$ cm$^{-2}$
and fit the spectra. The obtained values for $kT_{\rm e}$, log($\tau$),
and EM changed less than 5\%, 1\%, and 15\%, respectively. Values of
metal abundances changed less than 5\%.  All these changes were less
than the statistical errors. Therefore, such variation in the $N_{\rm
H}$ value does not affect our fitting results significantly.

\subsubsection{Reg-1}

In reg-1, we found that $kT_{\rm e}$ increased from the outer region (o)
toward the inner region (i) whereas log($\tau$) decreased. Values of EM
have a maximum just behind the shock front.  The radial distribution of
$kT_{\rm e}$ seen in figure~\ref{fig:ana:limb:kne} is consistent with
those shown in figure~\ref{fig:ana:limb:map}. The gradient is
qualitatively consistent with that expected from a model of simple blast
wave propagating into a homogeneous medium.

These results are qualitatively similar to those obtained by MTPK
whereas our results are systematically different from those of MTPK.
Comparing with those of MTPK, our results show 20$-$30\% higher values
of $kT_{\rm e}$, resulting in lower EM.  There are several reasons to
account for discrepancies between our results and those by MTPK.  One of
the major reasons is that we use the Masai model whereas MTPK used the
NEI model coded by J. Hughes (Hughes model; Hughes \& Helfand 1985,
Hughes \& Singh 1993). There are differences in the atomic codes used in
the Masai model (Kato model; Kato 1976) and those in the Hughes model
(RS model; Raymond \& Smith 1977).  Another reason is that the
calibration and understanding of the SIS as well as the XRT have been
greatly improved since the writing of MTPK. We employed the latest
calibration files, data processing software, and response matrices
compared with those of MTPK as explained in section~2. Therefore, we
believe that our current results are more reliable than those of MTPK.

Figure~\ref{fig:ana:limb:ab} shows the radial structures of the
abundances of heavy elements relative to cosmic values. We find
significant variations from (i) to (o) for O, Ne, Mg, and Fe.  The
radial profiles of all elements except Si show clear jump structure
between (l) and (m) which is located at 0.9 $R_{\rm s}$. Outside of 0.9
$R_{\rm s}$, all profiles show constant values. On the contrary,
interior of 0.9 $R_{\rm s}$, all profiles seem to show marginal
increases toward the inner region.

\begin{figure}[htbp]
 \centering
 \epsscale{.8}
	\plotone{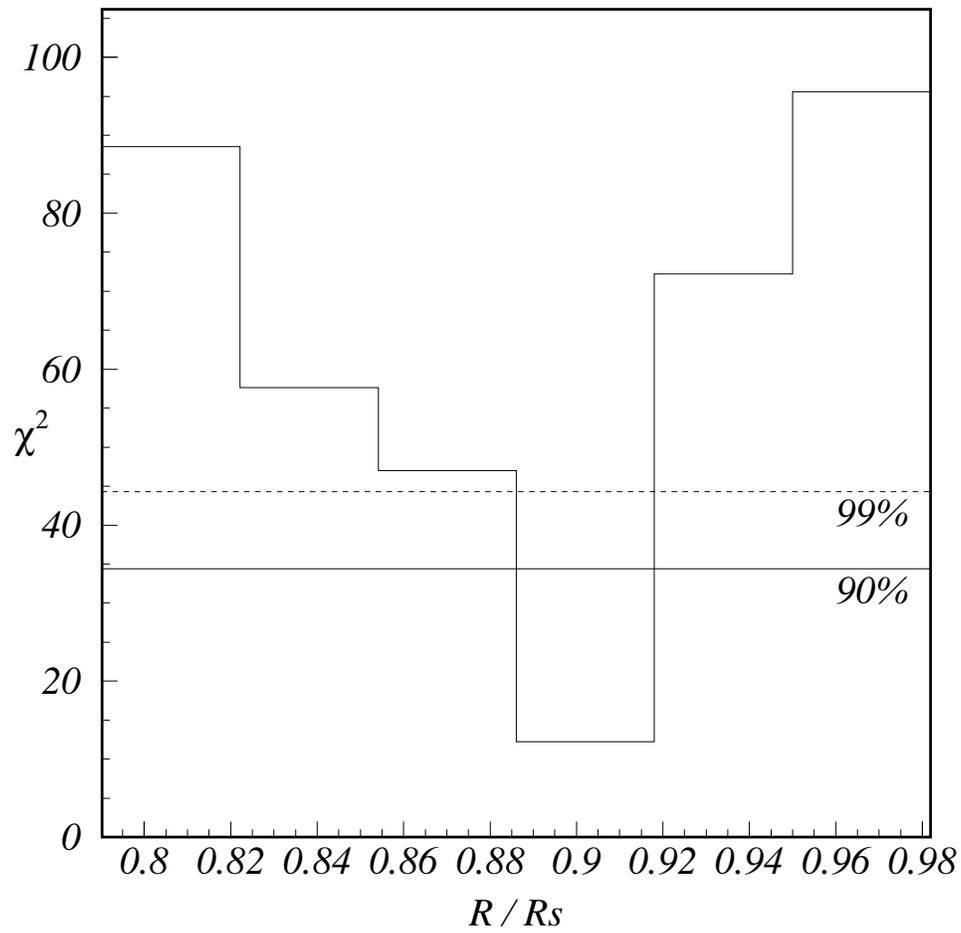}
 \caption{$\chi^2$ distributions for all elements fitted with the step
 function. Horizontal solid line shows 90 \% confidence level and dotted
 line shows 99 \% confidence level.}
 \label{fig:ana:limb:chi}
\end{figure}

We performed a statistical test on the radial distributions of metal
abundances. We assumed that abundances of all elements were constant
inside of $r_1$ ($0.75 < r_1 < 1$) and outside of $r_1$ they were also
constant but different values for all elements, where $r_1$ was a
projected radius. We fitted the abundance distributions for all
elements simultaneously as a function of $r_1$ based on this assumption.
Fitting result is shown in figure~\ref{fig:ana:limb:chi}.  Apparently,
the abundance distribution of heavy elements is not constant in
reg-1. There is only one acceptable radius of $r_1 \approx$ 0.9 $R_{\rm
s}$ even at 99 \% confidence level.  Therefore, the presence of the jump
structure in the radial distributions of heavy elements is statistically
significant.

\subsubsection{Reg-2}

In reg-2, the obtained values of $kT_{\rm e}$ are much higher than those
of reg-1 as shown in figure~\ref{fig:ana:limb:kne}. We should note that
$kT_{\rm e}$ values in the inner region of reg-2 are higher than those
in the center portion of the Cygnus Loop (Miyata et al. 1998). On the
contrary, values of log($\tau$) decrease toward the inner
region.

For the radial distribution of heavy elements shown in
figure~\ref{fig:ana:limb:ab}, there are no common tendencies although
scatter is large. The abundances of O and Ne at reg-2 are rather
smaller than those in reg-1 whereas that of Mg is similar to that of 
reg-1. Those of Si and Fe increase from reg-1 toward reg-2.

\section{Discussion}\label{sec:dis:limb}

\subsection{Radial Profiles of the Electron Gas}\label{sec:dis:limb:profile}

We first investigated radial profiles of physical parameters of reg-1
shown in figure~\ref{fig:ana:limb:kne}.  Our results reflect the
projected profiles along the line of sight. The theoretical calculations
usually describe unprojected profiles. Assuming the spherical symmetry,
we calculate the projected profiles to compare with our results.

We use two kinds of coordinate systems.  Both of them are centered on
the explosion center (${\rm \alpha = 20^h 49^m 15^s}$, ${\rm \delta =
30^\circ 51^\prime 30^{\prime\prime}}$ (1950); KKPL). One is the
observational system projected on the sky ($\boldmath r$). The other is
unprojected system of the SNR ($\boldmath R$).  In both systems, we use
the coordinate by normalizing the current location of the shock front,
$R_{\rm s}$.

\subsubsection{$kT_{\rm e}$ profile}\label{sec:dis:limb:kt}

\noindent
{\it (i) Sedov model}

\begin{figure}[htbp]
 \centering
 \epsscale{.8}
	\plotone{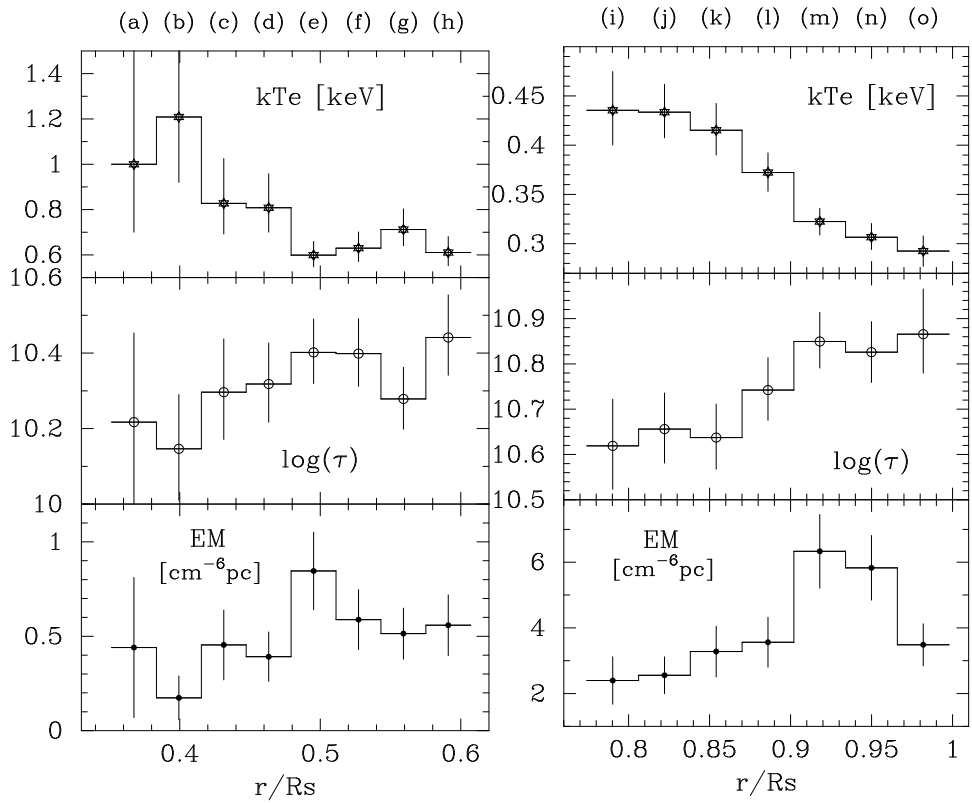}
 \caption{Radial profiles of $kT_{\rm e}$, log($\tau$), and EM.
 Horizontal axis shows the angular distance from the center
 normalized by the shock radius. Errors are at 90\% confidence level.}
 \label{fig:ana:limb:kne}
\end{figure}

\begin{figure}[htbp]
 \centering
	\plotone{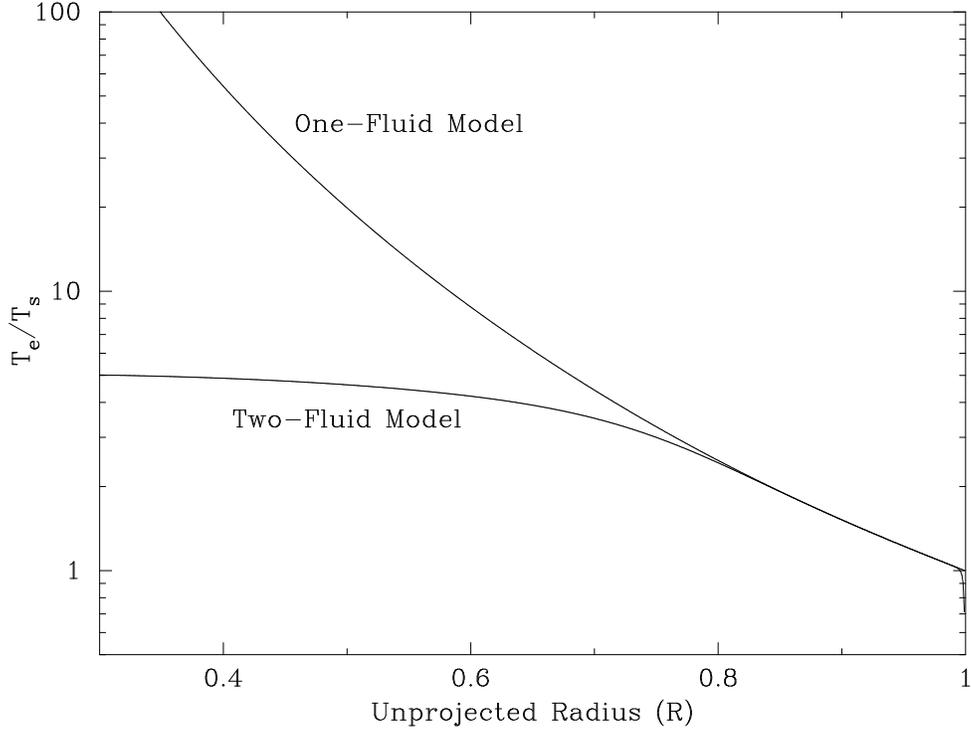}
 \caption{Unprojected radial profiles of $kT_{\rm e}$ based on the one-fluid
 and the two-fluid models. Parameters to be used for the two-fluid model are
 $E_{51}$ = 0.3, $t_3$ = 22, and $n_0$ = 0.2.}
 \label{fig:dis:limb:kt}
\end{figure}

As written in section~1, there are mainly two plausible models which
account for the electron heating mechanism: a one-fluid model and a
two-fluid model. Kahn (1975) developed an analytic method to describe
the physical structures in adiabatic SNRs based on the one-fluid
model. Cox \& Franco (1981) described the two-fluid condition by using
the approximation technique introduced by Itoh (1978) and derived an
approximation formulae to describe thermal structures for adiabatic
SNRs.  For the two-fluid model, the evolution of $kT_{\rm e}$ depends on
the explosion energy, $E_{51}$ (in units of \kaijo{10}{51} erg), the
density of the ambient medium, $n_0$ (in unit of ${\rm cm^{-3}}$), and
the age of the remnant, $t_3$ (in units of \kaijo{10}{3} yr).  We assume
$n_0$ to be $ 0.2\ {\rm cm}^{-3}$ (KKPL). $kT_{\rm e}$ just behind the
shock front ($T_{\rm s}$) is roughly 0.3~keV as shown in
figure~\ref{fig:ana:limb:map}. By using these parameters, we can
determine $E_{51}$ and $t_3$ with the Sedov equations (Sedov 1959),
\begin{eqnarray}
T_{\rm s } & =  & 1.5 \times 10^{10} \times \frac{E_{51}}{n_0}
 R_{\rm s}^{-3} \\
 & = & 1.2 \times 10^8 \ \frac{E_{51}}{n_0}^{2/5} t_3^{-6/5} \ ,
\end{eqnarray}
to be 0.3 and 22, respectively.  Applying these values, we calculate the
thermal structure based on the two-fluid model.
Figure~\ref{fig:dis:limb:kt} shows unprojected radial profiles of
$kT_{\rm e}$($R$) based both on the one-fluid model and on the two-fluid
model using the calculations by Cox \& Franco (1981).  As seen in this
figure, we expect that $kT_{\rm e}$($R$) increases toward the center
region in both cases. In the range of 0.8 $< R <$ 0.99, the
equipartition between ion and electron has achieved even if we consider
only Coulomb heating of electrons. Therefore, we expect that the radial
profile of $kT_{\rm e}$($r$) is similar if we use either the one-fluid
model or the two-fluid model. This figure shows the unprojected radial
profile of $kT_{\rm e}$($R$) while our result
(figure~\ref{fig:ana:limb:kne}) shows the projected profile.  We then
calculate the projected radial profile based on these models.

For simplicity, we assume that $kT_{\rm e}$ is determined by the
continuum emission through the spectral fitting procedure.  We thus
calculate the spectrum of the thermal bremsstrahlung emission
integrated along the line of sight weighted by EM for one-fluid
model and two-fluid model.  We adopt the Kahn's approximation for
$n_{\rm e}$($R$).  After calculating the projected spectra, we fit
them with the single $kT_{\rm e}$ thermal bremsstrahlung, considering
the effective area of SIS in the energy range of $0.4-4$ keV. Therefore,
the projected radial profile of $kT_{\rm e}$($r$) is valid only for the
results obtained with the SIS.

\begin{figure}[htbp]
 \centering
 \plotfiddle{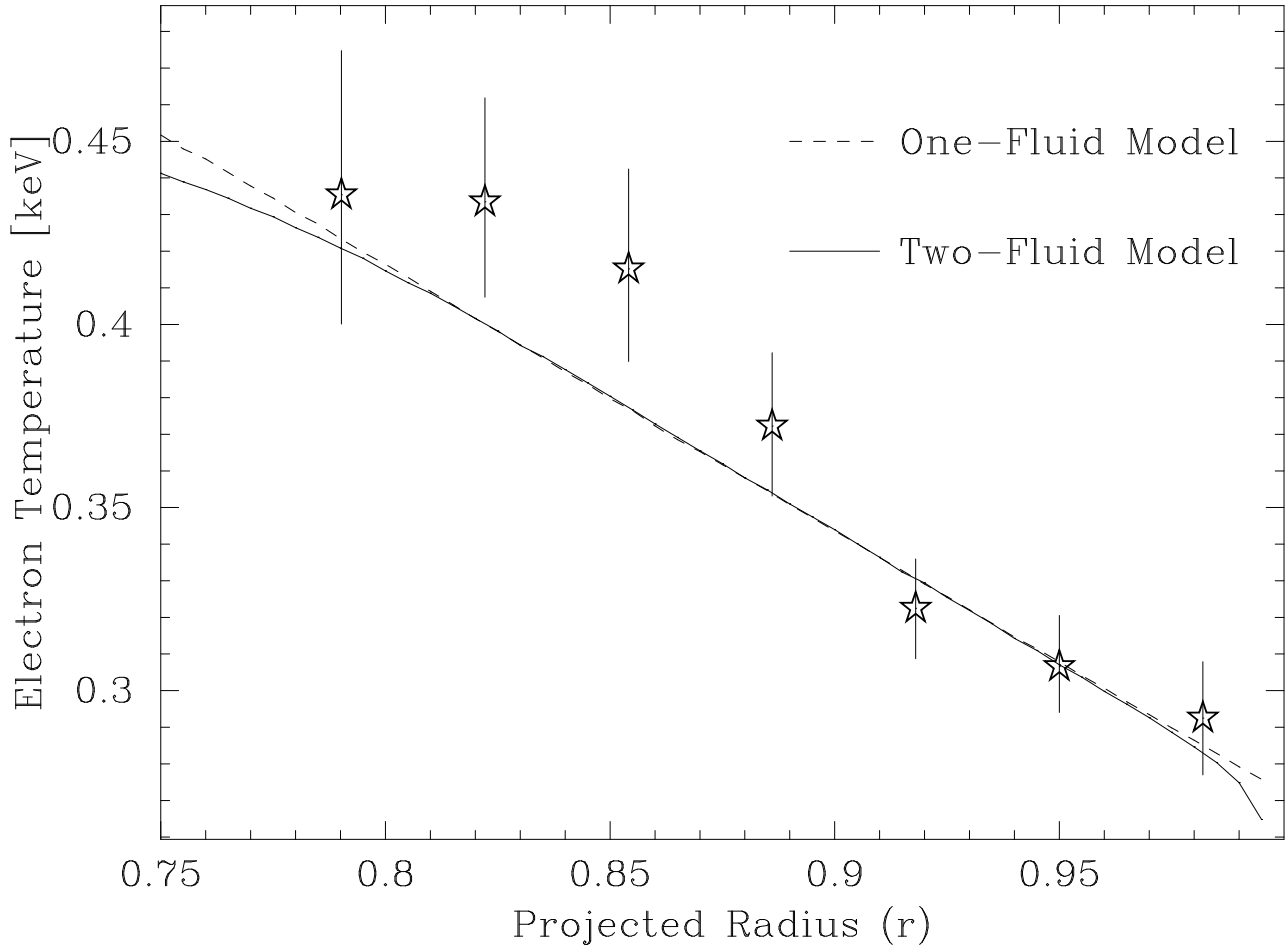}{13cm}{-90}{80}{80}{-300}{450}
 \caption{Projected radial profiles of $kT_{\rm e}$. Our results are shown
 by stars. Model calculations based on the one-fluid and the two-fluid 
 models are shown by dotted and solid lines,
 respectively.}
 \label{fig:dis:limb:kt_pro1}
\end{figure}

Figure~\ref{fig:dis:limb:kt_pro1} shows the projected radial profiles of
$kT_{\rm e}$($r$).  As we expected, the radial profiles of $kT_{\rm e}$($r$)
are similar for both models. Based on either model, $kT_{\rm e}$($r$)
linearly increases toward the center.  In the range of $ r >$ 0.75,
$kT_{\rm e}$($r$) is approximately described as ($ 0.98 - 0.71 \ r$)
keV. Our results show higher $kT_{\rm e}$ at $ r <$ 0.9 than those
expected from both models. We then modify the unprojected models to
reproduce our results.

\noindent
{\it (ii) Modified Two-Fluid Model}

\begin{figure}[htbp]
 \centering
 \plotone{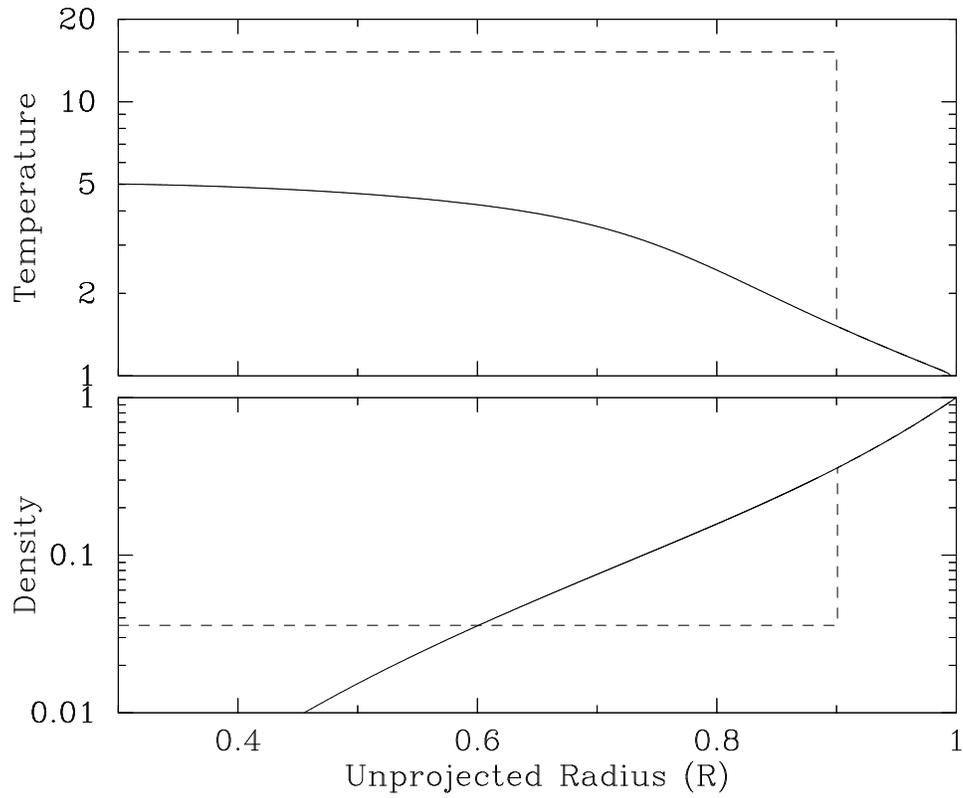}
 \caption{Unprojected radial profiles of $kT_{\rm e}$ 
 (upper pannel) and $n_{\rm e}$ (lower pannel), normalized
 by the values just behind the shock front.
 Solid lines and dashed lines show the radial profiles for the two-fluid
 model the modified two-fluid model, respectively. See text for details.}
 \label{fig:dis:limb:kt_dense_gap}
\end{figure}

As shown in figure~\ref{fig:ana:limb:ab}, the radial profiles of
abundances of heavy elements have a discontinuity at $r \approx$ 0.9. We
also see that the radial profile of $kT_{\rm e}$ as shown in
figure~\ref{fig:ana:limb:kne}, shows a jump structure at $r \approx $
0.9. Therefore, we calculate the radial profile of $kT_{\rm e}$($r$) in
the case that $kT_{\rm e}$($R$) has a discontinuity at $R$ = 0.9. In
this calculation, we adopt the two-fluid model since there is little
difference between one-fluid and two-fluid model as shown in
figure~\ref{fig:dis:limb:kt_pro1}.  In the region of $R \ge$ 0.9, the
radial profiles of $kT_{\rm e}$($R$) and $n_{\rm e}$($R$) are almost the
same as those by the two-fluid model as described in section
\ref{sec:dis:limb:kt}. For $R <$ 0.9, we introduce an extra parameter,
$\gamma$, to characterize the jump condition at $R$ = 0.9 in the radial
profiles of $kT_{\rm e}(R)$ and $n_{\rm e}$($R$). Here, we assumed that
the value of $kT_{\rm e}$ was constant at $R <$ 0.9 and that it was
$\gamma$ times higher than that of $kT_{\rm e}$ at $R$ = 0.9. For
pressure equilibrium, the value of $n_{\rm e}(R <$ 0.9) was assumed to
be $\gamma$ times lower than that of $n_{\rm e}$ at $R$ = 0.9. We call
this model a modified two-fluid model.
Figure~\ref{fig:dis:limb:kt_dense_gap} shows the radial profiles of
$kT_{\rm e}$ and $n_{\rm e}$ for the modified two-fluid model by dashed
lines as well as those for the two-fluid model shown by solid lines.
Here, we assumed $\gamma$ to be 10.  Values of $kT_{\rm e}$ and $n_{\rm
e}$ were normalized to those at $R_{\rm s}$.

\begin{figure}[htbp]
  \centering
     \plotfiddle{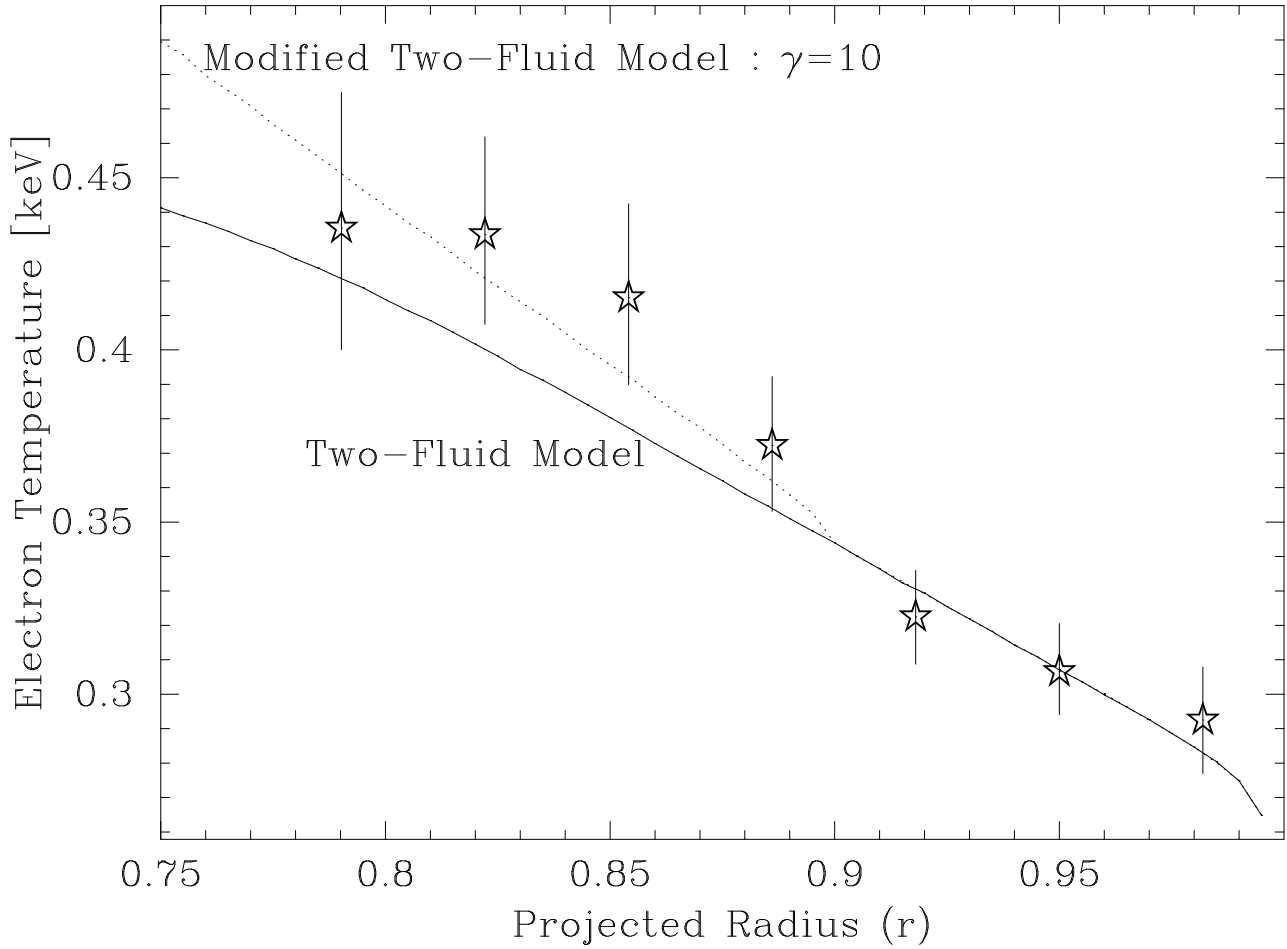}{13cm}{-90}{80}{80}{-300}{450}
 \caption{Projected radial profiles of $kT_{\rm e}$.
 Solid and  dashed lines are calculated
 based on the two-fluid model and the modified two-fluid model,
 respectively. See text for details.}
 \label{fig:dis:limb:kt_pro_gap}
\end{figure}

Based on this figure, we calculated the spectra at each unprojected
radius, integrated them along the line of sight, and fitted them at
each projected radius as the same way in section \ref{sec:dis:limb:kt}.
Results are shown in figure \ref{fig:dis:limb:kt_pro_gap}.

As mentioned in section \ref{sec:dis:limb:kt}{\it (i)}, values of $kT_{\rm
e}$(r) based both on the one-fluid model and on the two-fluid model are
lower than our results at $r <$ 0.9. For the modified two-fluid model,
the calculated profile was in reasonable agreement with our
data. Therefore, we preferred the modified two-fluid model rather than
one-fluid or two-fluid models without jump structures. Considering the
statistics of our data, $\gamma$ ranges between 5 and 15.

\noindent
{\it (iii) Radial Profiles from Reg-1 to Reg-2}

\begin{figure}[htbp]
 \centering
 \plotfiddle{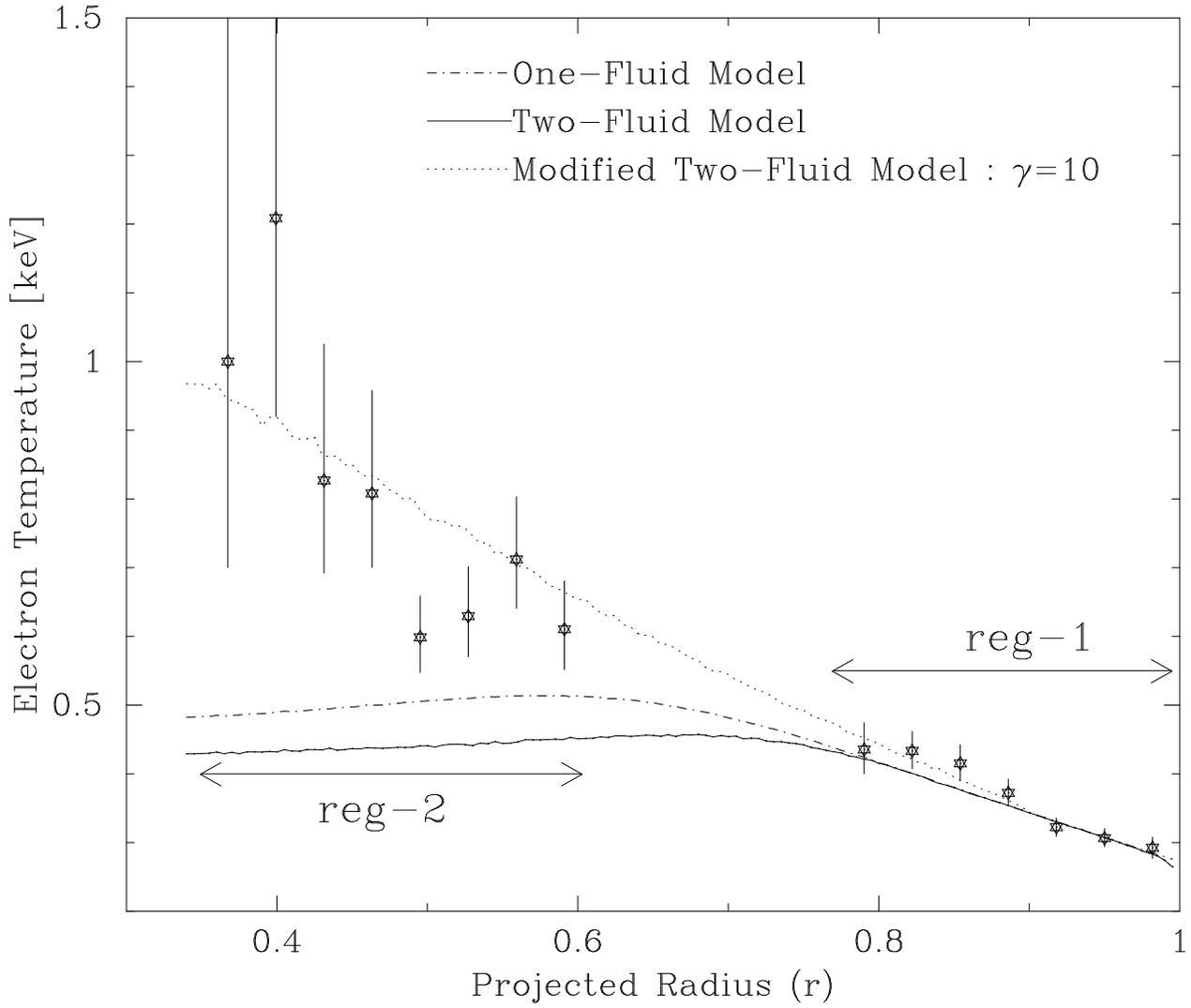}{13cm}{-90}{80}{80}{-300}{450}
 \caption{Same as figure~\ref{fig:dis:limb:kt_pro_gap}, but used data
 sets both reg-1 and  reg-2.}
 \label{fig:dis:limb:kt_pro_gap_withreg2}
\end{figure}

We then applied the modified two-fluid model both for reg-1 and for
reg-2. Figure~\ref{fig:dis:limb:kt_pro_gap_withreg2} shows the radial
profile of $kT_{\rm e}$ both for reg-1 and reg-2 as well as our
calculations based on the one-fluid model, the two-fluid model, and the
modified two-fluid model with $\gamma$=10. It is interesting to note
that apparent values of $kT_{\rm e}$ based on the one-fluid model
decrease toward the center at $r\leq$ 0.6. The expected $kT_{\rm e}$ at
the inner region is much higher than that at the shell region based on
the one-fluid model. Plasma with high $kT_{\rm e}$ ($\geq$ 10~keV) does
not affect our data while that with several keV can increase the
observed $kT_{\rm e}$ value.  In this sense, plasma at $R \geq$ 0.5 can
raise the observed $kT_{\rm e}$ only in the case that the emission
measure is enough to contribute.  As shown in
figure~\ref{fig:dis:limb:kt_dense_gap}, $n_{\rm e}$ at $R$=0.5 is
$\simeq$ 1 \% at $R$ = 1, resulting $\simeq$ 3 orders of magnitude lower
in emission measure. Such a small fraction is negligible and the
emission in the shell region is dominant in the calculation.  For this
reason, the calculated value of $kT_{\rm e}$ based on the one-fluid
model decreases toward the center.

The value of $kT_{\rm e}$ at $R \leq$ 0.9 based on the modified
two-fluid model is similar to that at $R \simeq$ 0.6 of the one-fluid
model. However, in the case of the modified two-fluid model, the
contribution from the hotter plasma becomes larger as one moves toward
the inner region.   Thus, $kT_{\rm
e}$ based on the modified two-fluid model increases toward the
center at $R\leq$0.6. Therefore, the modified two-fluid model with
$\gamma$ of 5--15 reasonably agrees with our results.  There are still
some deviations from our calculation at reg-2. This is probably due to
the simplicity of the model. More sophisticated model including some
patchy structure or local inhomogeneities would improve the fit to our
results.

Chevalier (1975a) initially pointed out an importance of the electron
thermal conduction for young SNRs and also suggested its importance on
the structure of the hot interior for evolved SNRs. Chevalier (1975b)
constructed an evolutionary model including the electron thermal
conduction based on the Mansfield \& Salpeter model (1974). Silk (1977)
also calculated the evolutional scenario including the electron thermal
conduction. In both models, $kT_{\rm e}$ goes up suddenly just behind
the shock front and stays constant value roughly at $R \leq$ 0.6. The
electron density has a sharp peak just behind the shock front due to a
shell formation, decreases from the shock front toward an inner region,
and stays constant value at $R \leq$ 0.6.  Chevalier's model and Silk's
model are similar to our model except for the location of the jump
structure.  Our observational result shows much stronger X-ray flux at
the shell region compared with their models. We suppose that the
constant $kT_{\rm e}$ at the inner region is due to the electron thermal
conduction.

\subsubsection{Emission Measure}\label{sec:dis:limb:em}

\begin{figure}[htbp]
 \centering
 \plotfiddle{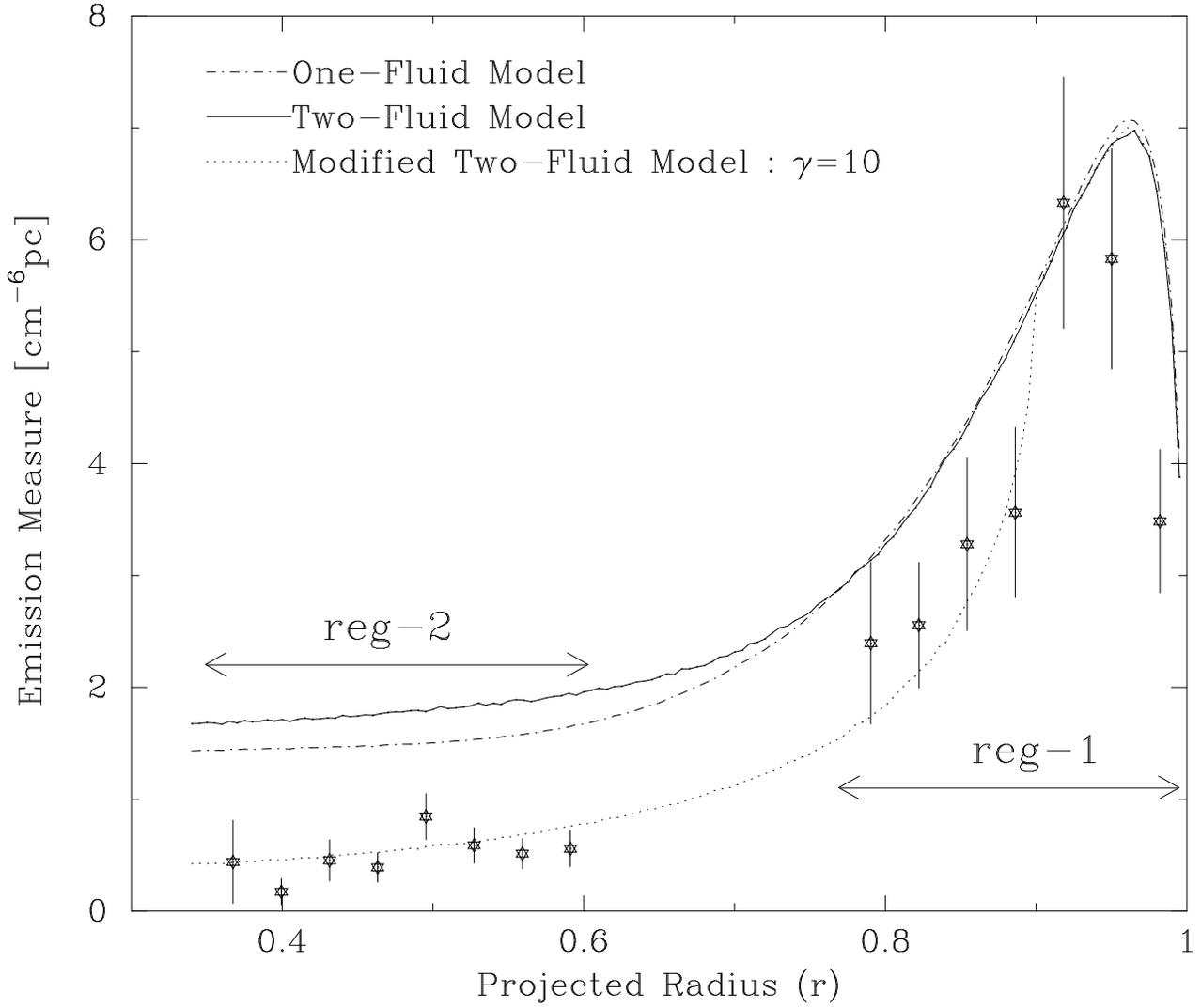}{13cm}{-90}{80}{80}{-300}{450}
 \caption{Radial profiles of EM for various
 models. Dash-dot, solid, and dotted curves are calculated
 based on the one-fluid model, the two-fluid model,
 and the modified two-fluid model.}
 \label{fig:dis:limb:variousEM}
\end{figure}

Next, we investigate the radial distribution of EM($r$) in the same way
as that of $kT_{\rm e}$($r$).  Figure~\ref{fig:dis:limb:variousEM} shows
EM($r$) for the three cases. Dash-dot, solid, and dotted lines show
EM($r$) based on the one-fluid, the two-fluid, and the modified
two-fluid model, respectively.

For the radial profile of EM($r$), neither the one-fluid model nor the
two-fluid model can reproduce our observational results inside of
$R\simeq$ 0.9. On the contrary, the modified two-fluid model with
$\gamma$ of 10 is in good agreement with our results. In this case, the
electron density of the ambient medium is roughly 0.25 ${\rm cm}^{-3}$.

Taking into account the result of the radial profile of $kT_{\rm e}$,
only the modified two-fluid model with $\gamma$ of 5--15 can reproduce
the SIS results.  The electron density of the inner region ($R\leq$ 0.9)
can be calculated as $\simeq 8 \times 10^{-3}{\rm cm}^{-3}$.  This
suggests that the hot tenuous plasma extends inside of $R\simeq$ 0.9.\\

\subsection{Jump in Metal Abundance at 0.9$R_{\rm s}$}

As shown in figure~\ref{fig:ana:limb:ab}, there is a jump structure at
$r\simeq 0.9$ in the radial distributions of heavy elements. Here, we
discuss the possible causes to explain the jump structure.

\subsubsection{Dust Sputtering ?}

The Copernicus satellite has studied the depletion along the line of
sight for various elements (see review for Spitzer \& Jenkins 1976). The
depletion of Fe compared with other elements like C, N, or O suggests
that Fe and other missing gas-phase elements are bound to interstellar
grains. After being engulfed by the shock wave, the dust grains
containing metal are destroyed in the postshock region due to the
nonthermal sputtering or due to grain-grain collisions (Seab \& Shull
1983). Employing the experimental data by KenKnight \& Wehner (1964) and
an approximation formula by Alm\'en \& Bruce (1961), the timescale of
the grain destruction, $t_{\rm s}$, is roughly estimated as,
\begin{equation}\label{eq:dis:limb:sputter}
  t_{\rm s} \,\,\approx \,\, 10^5 \,\,
	\left(\frac{n_{\rm e}}{1 {\rm cm}^{-3}}\right)^{-1}\,\,
	\left(\frac{T_{\rm e}}{10^6 {\rm K}}\right)^{-1/2}\,\,\,\,{\rm yr}.
\end{equation}
As dust grains are destroyed, heavy elements return to the gas phase.
Dust grains at the center regions would be sputtered for a longer time
than those at the shell, resulting the increase of abundances of heavy
elements toward the center. This picture seems to qualitatively explain
our results. However, as apparent in figure~\ref{fig:ana:limb:ab}, the
abundance of Ne increases toward the center. Since Ne is
a rare gas, there is no observational results of the depletion of Ne.
Therefore, it is difficult to explain the increase of abundances solely
by the dust sputtering model.

We should note that the derived Ne abundance contains some
ambiguity. Since emission lines from Ne overlap the energy range of Fe-L
emission line blends, the abundance of Ne is partly affected by that of
Fe.  There are also uncertainties in the atomic code of Fe-L blends
(e.g. Liedahl et al.  1995).  In the future, we will be able to detect
the Ne lines free from Fe-L blends to obtain the true Ne abundance using
X-ray detectors with a good energy resolving power.  To investigate the
model of the dust sputtering in detail, we need observations with higher
energy resolving power which can be realized in with the next Japanese
X-ray satellite, {\it Astro-E} (Inoue 1997).

\subsubsection{Contact Discontinuity ?}

A more straightforward way to explain the increase of abundances would
be with contributions of ejecta. Ejecta are considered to fill the region
inside of the contact discontinuity.  Based on Chevalier's similarity
solutions (1982), the location of the contact discontinuity ($R_{\rm
c}$) is at 0.73--0.90 $R_{\rm s}$ depending on density profiles of the
circumstellar matter and of the ejecta.  Since the ejecta fill
the region inside of the contact discontinuity, we expect an increase of
the abundances of heavy elements in the vicinity of the contact
discontinuity.  Based on figure~\ref{fig:ana:limb:ab}, we find a jump in
the radial profiles of abundances at $r \approx$ 0.9. This value agrees
well with the Chevalier's calculations in the case that the blast wave
propagates into a homogeneous medium. As Chevalier (1982b) pointed out,
however, $R_{\rm c}$ becomes smaller as SNR evolves into the Sedov
phase. During the Sedov phase, $R_{\rm c}$ is constant at $R\approx$
0.75 (McKee 1974), which is too far in compared with our results.
Therefore, it is not likely that the contact discontinuity exists at $R
= 0.9$.

\subsubsection{Shell Formation ?}

As described in section~\ref{sec:dis:limb:kt}, both Chevalier's model
and Silk's model suggested a sharp density peak just behind the shock
front due to a shell formation. This would be a signature that a SNR has
evolved into a radiative phase from an adiabatic Sedov phase. A width of
the density peak is expected to be $\simeq$ 1 \% of $R_{\rm s}$ which is
an order of magnitude narrower than the enhancement in emission measure
shown in figure \ref{fig:ana:limb:kne}. The expected $kT_{\rm e}$ is
also much lower than that can be detected with the ASCA Observatory.
Therefore, we ruled out that the density enhancement we observed was due
to the shell formation.

We should note that the Cygnus Loop has already shifted into radiative
phase at some places where we see optical filamentary structures
(e.g. Raymond et al. 1988). Such filamentary structures trace radiative
shock fronts. In our region, Hester et al. (1994) reported
Balmer-dominated filaments due to nonradiative shocks as well as the
H$\alpha$ filaments due to radiative shocks. Therefore, the radiative
shock must have taken place in the NE region of the Cygnus Loop.
However, we could not detect the density enhancement expected from the
radiative shock due to the poor spatial resolving power. Future X-ray
observations possessing much higher spatial resolution, like {\it CXO}
and {\it XMM}, will clarify the shock structure in the NE region.

\subsubsection{Pre-existing cavity ?}

We found that the density of the bright shell region was much higher
than that predicted by the model that the blast wave expanded into the
homogeneous medium as mentioned in
section~\ref{sec:dis:limb:profile}. On the contrary, the density inside
the shell was constant and quite low ($\simeq 8 \times 10^{-3}{\rm
cm}^{-3}$). These results are naturally interpreted with the hypothesis
that the supernova (SN) which produced the Cygnus Loop occured within a
pre-existing cavity wall (which was probably produced by the precursor).
This hypothesis has been widely supported by infrared, optical, and
X-ray observations.  If a cavity exists before the SN explosion, the
blast wave freely expands into the low density medium
and suddenly hits the high density
pre-existing cavity wall.  The kinetic energy of the blast wave can be
converted to thermal energy in the cavity and heat it up to emit
X-rays. As the result, we expect that the metal abundances at the bright
shell region reflect those of the ISM while the metal abundances at low
density inner region reflect those of ejecta. This scenario fits our
results both with respect to the radial profiles of $kT_{\rm e}$,
EM, and the abundances of heavy elements.

\subsection{Radial Profile in Metal Abundances}

Miyata et al. (1998) found Si, S, and Fe rich plasma at the center
portion of the Cygnus Loop. Such plasma was confined in a small region
with the radius of $\simeq$ 9 \arcmin. Our results suggest that Si and Fe
may increase from (d) toward the center. Therefore, the major part of Si
and Fe in ejecta might be confined well within the shell region.

If the ejecta fill only inside of 0.9 $R_{\rm s}$, the abundances in
(m)--(o) reflect the ISM alone or the ISM contaminated by stellar wind
from the progenitor star.  Abundances of O and Ne at reg-2 are similar
to those of reg-1 ($r < 0.9$). These values are also similar to those
obtained in the center portion of the Cygnus Loop (Miyata et al. 1988).
Therefore, O and Ne in ejecta are distributed inside of $R_{\rm s} \leq
0.9$.  Mg abundances in reg-2 are similar to those in (m)--(o),
suggesting that the Mg in the ejecta is distributed possibly in $0.6
\leq R_{\rm s} \leq 0.9$.

Mg abundance is distributed in the outer region compared with those of
Si and Fe while O and Ne are uniformly distributed inside the shell.
Based on the nucleosynthesis model in supernova calculated by Thielemann
et al. (1996), heavy elements like Si, S, and Fe are produced in the
inner region whereas light elements like O, Ne, and Mg are produced in
the outer region of the progenitor star. Some fraction of Si and S are
also synthesized in the outer region where O is the dominant
element. This picture is in good agreement with our observational
results.  Therefore, our result is a possible evidence of the onion-skin
structure in such an evolved SNR as the Cygnus Loop.

\subsection{Progenitor Mass}

Assuming $\gamma$ to be 10 in the modified two-fluid model, we can
calculate the total mass contained inside $R$ = 0.9. The volume and the
density inside $R$ = 0.9 ($\approx$ 16.6~pc) are $1.9 \times 10^4 {\rm
pc}^3$ and $8\times 10^{-3}{\rm cm}^{-3}$, respectively. The total
mass is roughly 4\Msun. If this is the progenitor mass, its
main-sequence mass is estimated to be \kara 15\Msun (Nomoto, Hashimoto
1984).  This strongly supports the massive star origin of the Cygnus
Loop.  

Charles et al. (1984) suggested the spectral type of the progenitor star
of the Cygnus Loop to be later than B0 based on the radius of the
cavity.  Shull \& Hippelein (1991) restricted the spectral type to be B
stars rather than O stars. Since O stars tend to destroy almost anything
around them due to the photoevaporation, they would not have left
well-defined cavities or shells (Shull et al. 1985). Chevalier (1988)
pointed out that 15\Msun\ was a critical mass above which the
circumstellar effects were large and below which they disappeared
rapidly. This is due to the fact that the rate of emission of ionizing
photons drops dramaticaly for stars later than B0 stars (Panagia
1973). Therefore, our estimate of 15\Msun\ is consistent with the view
of the pre-existing cavity.

In this case, the H-rich envelope mass of the progenitor star is roughly
10\Msun. If it is present outside of 0.9 $R_{\rm s}$, the width of the
H-rich envelope is estimated to be \kara 0.3 pc, which is \kara
1\arcmin. This cannot be resolved with the ASCA Observatory. The {\it
XMM} may reveal the plasma structure in the shell region in detail.

\subsection{Mixing of the Ejecta with the ISM}

\begin{figure}[htbp]
 \centering
 \epsscale{.8}
	\plotone{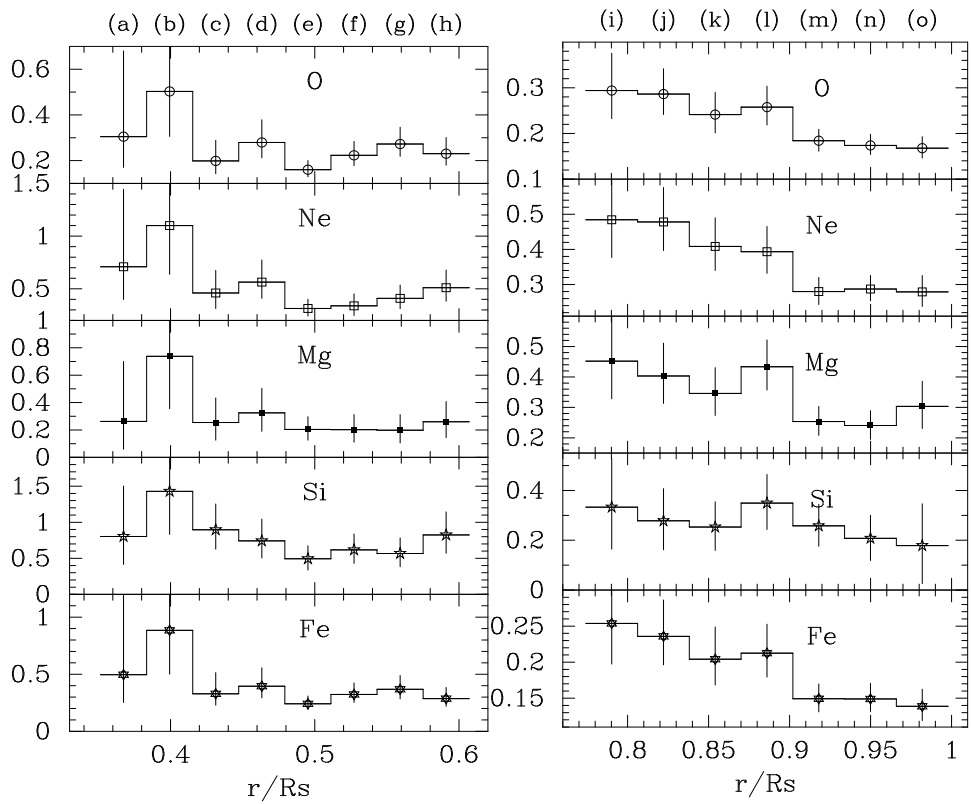}
 \caption{Same as figure~\ref{fig:ana:limb:kne} but for
 abundances of heavy elements relative to cosmic values.}
 \label{fig:ana:limb:ab}
\end{figure}

We estimate the timescale for mixing of the ejecta with the shocked
pre-existing cavity. The diffusion timescale, $\tau_d$, for the
ejecta to mix with the surrounding gas is approximately
\begin{equation}\label{eq:dis:limb:diffusion}
 \tau_d = \frac{L^2}{D}\ \ ,
\end{equation}
where $L$ is the typical size of the mixing region and $D$ is the diffusion
coefficient. Taking into account Coulomb collisions solely, the mean
free path of an ion, $\lambda$, is
\begin{eqnarray}\label{eq:review:mfp}
  \lambda &=& t_{\rm ee} \left(\frac{3 kT_{\rm e}}{m_{\rm e}}\right)^{1/2} \\
        &=& 0.29 \ \left(\frac{T_{\rm e}}{10^7 {\rm K}}\right)^2
          \left(\frac{n_{\rm e}}{1 {\rm cm}^{-3}}\right)^{-1}
          \left(\frac{\ln\Lambda}{30}\right)^{-1}
          \ \ \ {\rm pc} \ \ \ ,
\end{eqnarray}
where $t_{\rm ee}$ is the relaxation timescale of the electron gas
and $\ln\Lambda$ is the Coulomb logarithm.
Then, $D$ is (Chevalier 1975)
\begin{equation}
 D = \frac{1}{3} \lambda v
  = 1.0 \times \kaijo{10}{23}
  \left(\frac{T_{\rm i}}{5 \times 10^6 {\rm K}}\right)^{5/2}
  \left(\frac{n_{\rm i}}{0.1 {\rm cm}^{-3}}\right)^{-1}
  \left(\frac{Z}{10}\right)^{-2}
  \ \ \ {\rm cm}^2\ {\rm s}^{-1} \ \ \ ,
\end{equation}
where $T_{\rm i}$ is the ion temperature and $n_{\rm i}$ is the ion
density. $Z$ is the mean charge of heavy elements and roughly 10 in our
case (Masai 1984).  Since we find the boundary between the
ejecta and the cavity as shown in figure~\ref{fig:ana:limb:ab}, the
transition region is smaller than the size of each annular sector which
is equal to the PSF of the XRT. So we set $L$ to be $\le$ 1~pc. The
timescale for the mixing is
\begin{equation}
 \tau_d = 3.0 \times 10^6
  \left(\frac{L}{1~{\rm pc}}\right)^2
  \left(\frac{T_{\rm i}}{5 \times 10^6 {\rm K}}\right)^{-5/2}
  \left(\frac{n_{\rm i}}{0.1 {\rm cm}^{-3}}\right)
  \left(\frac{Z}{10}\right)^{2} \ \ \ {\rm yr} \ \ \ .
\end{equation}
As mentioned by Chevalier (1975), this estimate may be shortened by the
turbulent diffusion whereas this may be lengthened by the work of the
magnetic field.  However, these effects have not yet been
established. The timescale shown above is a first approximation for
the mixing of the ejecta in the shocked pre-existing cavity. Apparently,
this timescale is much longer than the age of the Cygnus
Loop. Therefore, the ejecta will not mix with the pre-existing cavity
for next $10^6$ yr.

\acknowledgments

We are grateful to all the other members of the ASCA team. We thank to
Dr. Shigeyama for useful discussions and suggestions.  Dr. B. Aschenbach
kindly gave us the entire X-ray image of the Cygnus Loop obtained with
the ROSAT all-sky survey.  We would like to thank the anonymous referee
for her or his detailed comments and suggestions, which greatly improved
this paper. EM is partially supported by Inoue Research Award for Young
Scientists.

\clearpage

\clearpage

\clearpage

\clearpage

\clearpage

\clearpage

\end{document}